\documentclass[aps,pra,twocolumn,showpacs]{revtex4}
\usepackage{amssymb}
\usepackage{color}
\usepackage{amsmath}
\usepackage{graphicx}
\usepackage{bm}
\def\dbar{{\mathchar'26\mkern-12mu d}} 
\begin{document}

\title{Quantum Otto engine with a spin $1/2$ coupled to an arbitrary spin}


\author{Ferdi Altintas}\email{ferdialtintas@ibu.edu.tr}

\affiliation{Department of Physics, Abant Izzet Baysal University, Bolu, 14280, Turkey}

\author{\"{O}zg\"{u}r E. M\"{u}stecapl{\i}o\u{g}lu}
\email{omustecap@ku.edu.tr}
\affiliation{Department of Physics, Ko\c{c} University, \.{I}stanbul, Sar{\i}yer, 34450, Turkey}

\begin{abstract}
We investigate a quantum heat engine with a working substance of two particles, one with a spin $1/2$ and the other with an arbitrary spin (spin $s$), coupled by Heisenberg exchange interaction, and subject to an external magnetic field. The engine operates in a quantum Otto cycle. Work harvested in the cycle and its efficiency are calculated using quantum thermodynamical definitions.  It is found that the engine has higher efficiencies at higher spins and can harvest work at higher exchange interaction strengths.The role of exchange 
coupling and spin $s$ on the work output and the thermal efficiency is studied in detail. In addition, the engine operation is analyzed from the perspective of local work and efficiency. The local work definition is generalized for the global changes and the conditions when the global work can be equal or more than the sum of the local works are determined.
\end{abstract}

\maketitle
\section{Introduction}\label{sec:intro}
The investigations of heat engines in the quantum regime, or quantum thermodynamics, has become an active area of research in the last decade~\cite{scovil59,quan07,quan09,kieu04,jwang12,wang13,uzdin14,quan05,allahverdyan08,altintas14,thomas11,zhang08,huang13,
feldmann04,feldmann03,kosloff02,henrich07,zhang07,thomas14,huang12,huang14,wu06,ivanchenko14,wang12,he12,
huang14plus,wang09,hübner14,azimi14,albayrak13,dillen09,rezek06,robnagel14,zhang14,scully03,quan06,altintas15,
sothmann12,sothmann14,fialko12,zhangprl14,abah12}.  A quantum heat engine (QHE) uses a quantum working substance 
to harvest work in a quantum thermodynamical cycle~\cite{quan07,quan09,kieu04}. 
Three level masers can be considered as the first  QHEs~\cite{scovil59}. Prototype quantum systems, such as two level ~\cite{quan07,quan09,kieu04,jwang12,wang13} and multilevel particles~\cite{uzdin14,quan05,allahverdyan08}, coupled spins~\cite{altintas14,thomas11,zhang08,huang13,feldmann04,feldmann03,kosloff02,henrich07,zhang07,thomas14,huang12,huang14,
wu06,ivanchenko14,wang12,he12,huang14plus,wang09,hübner14,azimi14,albayrak13}, and harmonic oscillators~\cite{dillen09,rezek06,robnagel14,zhang14} are considered as quantum working substances. Circuit and cavity quantum electrodynamics systems~\cite{scully03,quan06,altintas15}, quantum dots~\cite{sothmann12}, quantum Hall 
edge states~\cite{sothmann14}, cold bosonic atoms~\cite{fialko12}, optomechanical systems~\cite{zhangprl14}, and a 
single ion~\cite{robnagel14,abah12} 
have been proposed  to realize QHEs; while ultracold atoms are porposed for work measurements~\cite{roncaglia14}. 
In addition to the studies focusing on the quantum properties, such as 
quantum coherence and correlations, of the working substance~\cite{altintas14,zhang08,zhang07,huang12,wang12,he12,wang09,hübner14,azimi14,albayrak13}, there are explorations of the quantum
heat reservoirs as well~\cite{huang12,robnagel14,zhang14,scully03,quan06}. 

In the present contribution, we assume classical heat reservoirs, and consider two interacting particles, one with a spin $1/2$ and the other with an arbitrary spin (spin $s$), as our working medium. 
The particles are assumed to be in an external magnetic field and they interact with each other by Heisenberg exchange coupling. The two spin $1/2$ case of this model has been a subject of much attention in quantum thermodynamics ~\cite{altintas14,thomas11,zhang08,huang13,feldmann04,feldmann03,kosloff02,henrich07,zhang07,thomas14,
huang12,huang14,wu06,ivanchenko14,wang12,he12,huang14plus,wang09,albayrak13}.  An appealing property of the Heisenberg model is that the quantum Otto engine efficiency
can be enhanced at a critical exchange interaction between two spins $1/2$~\cite{thomas11}. We consider the arbitrary spin $s$ as another control parameter next to the exchange coupling and explore its influence on the performance of the QHE. Such higher spin Heisenberg models could be implemented for QHE operations in nuclear magnetic resonance (NMR) systems~\cite{sinha03}.
Among typical quantum thermodynamical cycles~\cite{quan07,quan09} we choose to operate our QHE in the Otto cycle as it consists of less demanding processes to implement  in comparison to other quantum cycles and proposed in various systems for implementations~\cite{robnagel14,fialko12,abah12,altintas15}. 
 
The system we consider can be interpreted as the central spin (Gaudin) model with homogenous couplings~\cite{bortz07}. The large spin $s$ in our model plays the role of a collective spin bath~\cite{prokof00} consisting of $2s$ spins $1/2$, which are homogeneously coupled to a central spin $1/2$. Recent studies revealed that a spin $1/2$ ensemble can be used as a heat reservoir to a single spin $1/2$, if it consists of at least two spins 1/2~\cite{klein15}. In our case there is an additional heat reservoir coupled both to the central spin and to the collective spin bath. In such a case, the central spin 1/2 can always be thermalized  while spin $s$ cannot if $s>1/2$. Even when spin $s$ is not in thermal equilibrium, the total system is always fully thermalized. 
The coupled spin $1/2$ and spin $s$ model is hence far from a trivial extension of coupled spin $1/2$ system but an intriguing generalization. 

Consequences of the coupling heat and work reservoirs to interacting asymmetric spins lead to surprising results which cannot be expected and understood by the knowledge accumulated from the models of coupled spins $1/2$. For example, we calculated the work output and efficiency of our model QHE. Our results show that with an arbitrary spin $s$, one can extract more work with higher efficiency than the two spins 1/2 case given in Ref~\cite{thomas11}. Especially the upper bound of efficiency for the two spins $1/2$ given in Ref.~\cite{thomas11} can be beaten by an arbitrary spin $s$. In addition, local and
global thermodynamics of asymmetric spins exhibit peculiar differences from those of two spins $1/2$. Asymetric two spins with $s>1/2$ can act as a QHE even in the ultrastrong coupling regime, contrary to the two spins 1/2. Furthermore, spin $1/2$ can be a refrigerant in this regime, while spin s dominates the QHE behavior for the total system. Local thermodynamics can be explored deeper by the concepts of local heat and temperature. Surprisingly local temperature is not applicable to spin $s>1/2$ when it is coupled to the spin $1/2$. The local temperature of spin $1/2$ is always well defined. It can be controlled by spin $s$
and can be made negative or larger than the temperature of the heat reservoir. 

In addition, we introduce generalized definitions of local and cooperative work to explore local thermodynamics even when global parameters in a bipartite system varies. The earlier definitions are limited to the variation of local parameters only. Our formalism explicitly relates the quantum covariance, or quantum fluctuations and quantum coherence, to the work output of QHE 
of any interacting bipartite system. Within the developed framework, conditions for the violation of the extensive behavior of the global work as well as a measure of quantum cooperativity in terms of covariance naturally emerge. While the formalism is model independent,
we provide an illustration by applying it to the asymmetric spin system.

The paper is organized as follows. In Sec.~\ref{sec:QHEmodel}, we introduce our model QHE. The results for the global and local engine operations are given in Secs.~\ref{sec:globalWorkEfficiency} and~\ref{sec:LocalWorkEfficiency}, respectively. A general discussion on
the relation between global and local work is given in Sec.~\ref{sec:generalGlobalLocalWork}. The conclusions are stated in Sec.~\ref{sec:conclusion}.
\section{Model quantum heat engine}\label{sec:QHEmodel}
The working substance of our QHE consists of two spins in a homogeneous magnetic field, coupled to each other with a 
Heisenberg exchange interaction and it is described by a Hamiltonian~\cite{thomas11,ivanchenko14,li12}:
\begin{eqnarray}\label{hamiltonian}
H=8J\vec{s}_A.\vec{S}_B+2B\left(s_A^z+S_B^z\right),
\end{eqnarray}
where $\hbar=1$ is taken. $\vec{s}_A=\left(s_A^x,s_A^y,s_A^z\right)$, $\vec{S}_B=\left(S_B^x,S_B^y,S_B^z\right)$, $s_A^i$ and $S_B^{i}$ $(i=x,y,z)$ are the spin $1/2$ and spin $s$ operators, respectively. Here, we label the spin $1/2$ and spin $s$  
with $A$ and $B$, respectively. The factor $B$ in the second term of the Hamiltonian denotes the external homogeneous
 magnetic field applied along the $z$ axis. 
We take $\mu_B=1$ and assume there is no orbital angular
momentum so that the gyromagnetic ratio $\gamma$ is the same for both spins, $\gamma=2$. $J~(\geq 0)$ is the anti-ferromagnetic coupling constant. Here we restrict ourselves to $s=1/2,1,3/2,2,5/2,3$.

The eigenvalues $E_n$ of the model Hamiltonian are tabulated in Appendix. In thermal equilibrium with a heat bath at temperature $T$ 
the density matrix $\rho$ of the working medium can be written as
\begin{eqnarray}\label{denmatAB}
\rho=\sum_nP_n\left|\Psi_n\right\rangle\left\langle\Psi_n\right|.
\end{eqnarray}
The occupation probabilities of the eigenstates $\left|\Psi_n\right\rangle$ are $P_n=\exp{(-E_n/T)}/Z$ ($k_{B}=1$) and 
$Z=\sum_n\exp{(-E_n/T)}$ is the partition function.

We consider the working medium described by the Hamiltonian in Eq.~(\ref{hamiltonian}) undergoes a quantum Otto cycle which consists of two quantum adiabatic and two quantum isochoric processes. The adiabatic branches involve the change of magnetic field between two chosen values $\left(B_1\rightarrow B_2\rightarrow B_1\right)$ at a fixed coupling strength, $J$. The details of the cycle are described below.

{\it Stage 1.} This stage is the {\it quantum isochoric process}, where the working medium with external magnetic field $B_1$ and coupling constant $J$ interacts with a heat bath at $T=T_1$. The interaction takes long enough, so that the working substance falls into a steady state given by Eq.~(\ref{denmatAB}) with occupation probabilities $P_n$ and energy levels $E_n$. {\it Stage 2.} The working medium undergoes 
{\it a quantum adiabatic process}, in which the interaction between the system and the heat bath is turned off and the magnetic field is changed from $B_1$ to $B_2$. The quantum adiabatic theorem is considered to hold (provided the process is slow enough)~\cite{thomas14}, so that the occupation probabilities remain unchanged, while the energy levels change from $E_n$ to $E_n^{'}$ due to the change in the magnetic strength. {\it Stage 3.} This process is almost the reverse of Stage 1, where the working medium is in contact with a cold heat bath at $T=T_2$ $(T_1>T_2)$. Reaching equilibrium with the bath changes the energy probabilities to $P_n^{'}$ with $B=B_2$, $T=T_2$ and $J$ in Eq.~(\ref{denmatAB}). {\it Stage 4.} The system undergoes another quantum adiabatic process with changing $B_2$ to $B_1$ ($E_n^{'}$ to $E_n$), while keeping $P_n^{'}$ the same. 

From the generalization of the first law of thermodynamics to quantum mechanical systems~\cite{quan07,quan09,kieu04}, the heat exchanges in Stages 1 and 3 are, respectively, given as
\begin{eqnarray}\label{heatAB}
Q_1&=&\sum_nE_n\left(P_n-P_n^{'}\right),\nonumber\\
Q_2&=&\sum_nE_n^{'}\left(P_n^{'}-P_n\right).
\end{eqnarray}
The work is performed only in the adiabatic branches of the quantum Otto cycle. Due to the conservation of energy, the net work done by the QHE can be written as:
\begin{eqnarray}\label{workAB}
W&=&Q_1+Q_2\nonumber\\
&=&\sum_n\left(E_n-E_n^{'}\right)\left(P_n-P_n^{'}\right),
\end{eqnarray}
where $W>0$ signifies the work performed by the QHE with operational efficiency $\eta=W/Q_1$. To harvest positive work by the engine, we consider $Q_1>-Q_2>0$ to conform to the second law of thermodynamics.

By using the tabulated eigenvalues $E_n$ of $H$ in Appendix and the probabilities given by the thermal occupation numbers in Eq.~(\ref{denmatAB}), the work output and the efficiency of the engine can be calculated analytically. The analytical expressions are 
not very illuminating and will not be displayed here for brevity. We call the work done by the engine given by Eq.~(\ref{workAB}) and its efficiency  $\eta$ as the global work and global efficiency, respectively, to distinguish them from the local work and efficiency of individual spins, described later in the text.
\section{Global Work and Efficiency}\label{sec:globalWorkEfficiency}
Before presenting our results, we would like to review some of the main results in Ref.~\cite{thomas11} where the authors investigated the same Hamiltonian in Eq.~(\ref{hamiltonian}) but for two spins $1/2$. The conditions in which the coupled engine efficiency can be higher than the uncoupled one have been determined. Specifically, an upper bound $\eta_b$ to the efficiency $\eta$ 
of the quantum Otto engine has been obtained as
\begin{eqnarray}\label{bound}
\eta\leq\eta_b=\frac{1-B_2/B_1}{1-4J/B_1}<\eta_c,
\end{eqnarray} 
where the upper bound is always less than the classical Carnot efficiency $\left(\eta_c=1-T_2/T_1\right)$.

\begin{figure}[!ht]\centering
\includegraphics[width=8cm]{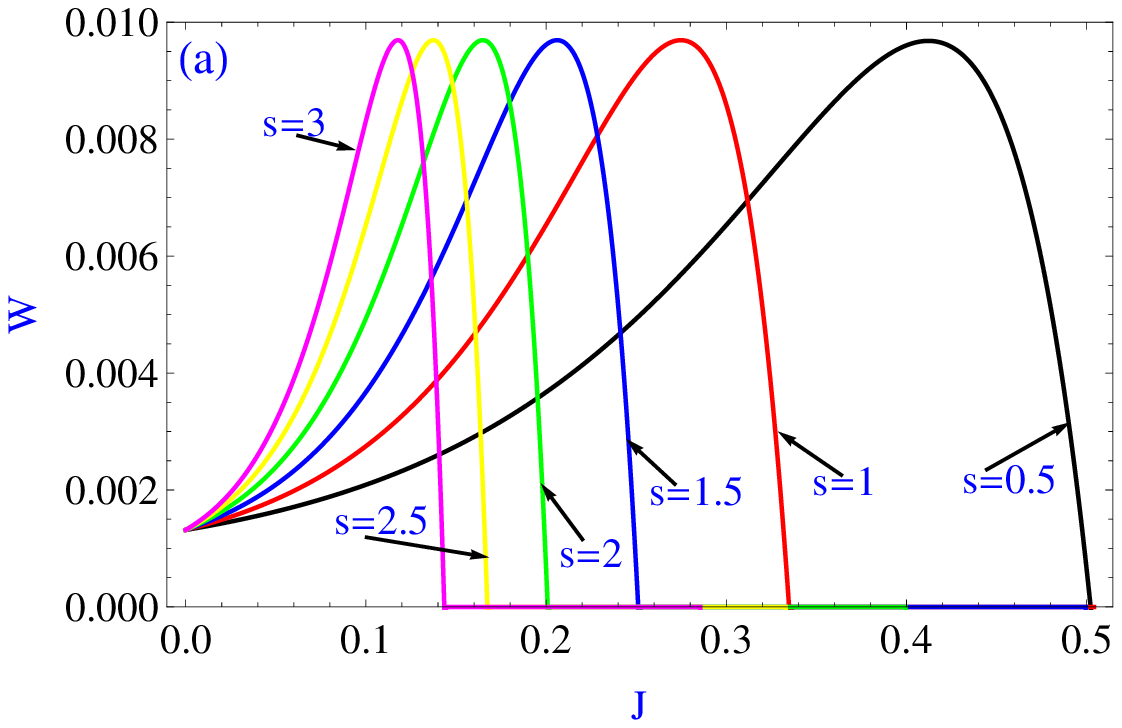}
\includegraphics[width=8cm]{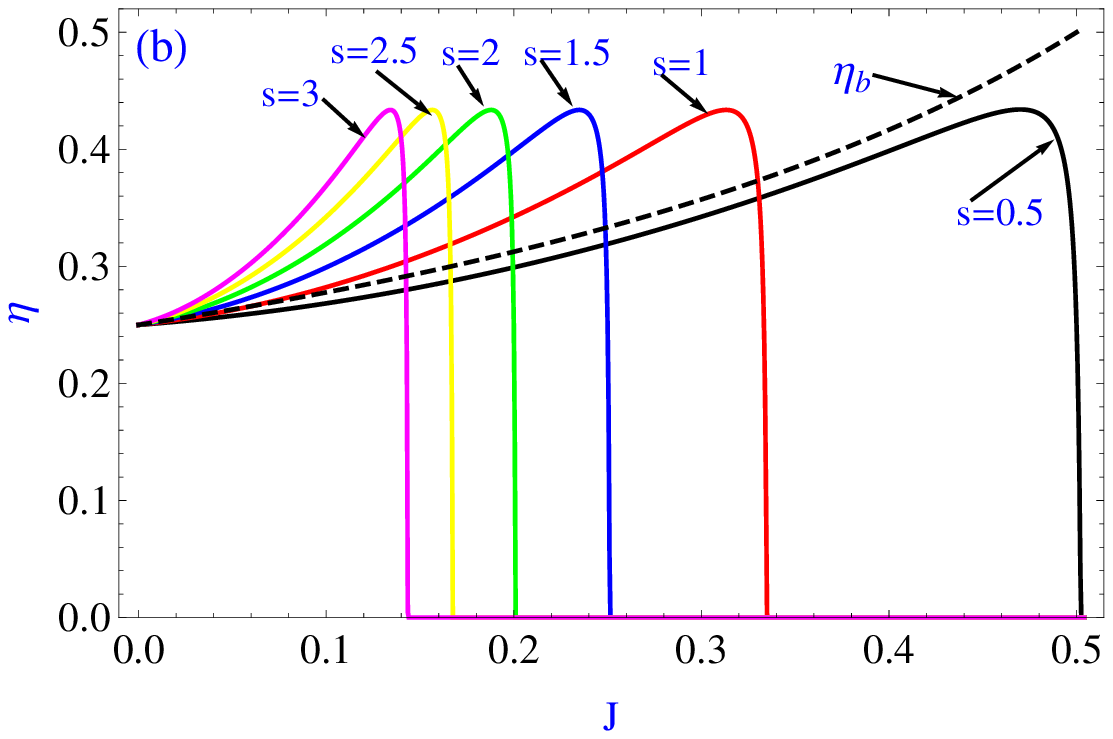}
\caption{\label{fig:fig1}(Color online.) Dependence of global work $W$ (a) and efficiency $\eta$ (b) on coupling strength $J$ for temperatures $T_1=1$, $T_2=0.5$, and magnetic fields $B_1=4$, $B_2=3$, and spins $s=1/2$~(black line), and $s=1$~(red line), $s=3/2$~(blue line), $s=2$~(green line), $s=5/2$~(yellow line) and $s=3$~(magenta line). The dashed line in (b) indicates the upper bound $\eta_b$ of the global efficiency given in Eq.~(\ref{bound}) for the case of spin $1/2$ pair. For the above parameters, we have $\eta_c=1-T_2/T_1=0.5$ and $\eta_{J=0}=1-B_2/B_1=0.25$. All quantities plotted are dimensionless. In all figures, we use a unit system where $\hbar=1,\mu_B=1,k_B=1$ and use 
$T_1$ as our scaling parameter.}
\end{figure}
In Fig.~\ref{fig:fig1}, we investigate the role of spin $s$ on the performance of the coupled quantum Otto engine. We plot the global work in 
Fig.~\ref{fig:fig1}(a) and global efficiency in Fig.~\ref{fig:fig1}(b), as a function of exchange coupling strength $J$ for 
$B_1>B_2$ and $s=1/2,1,3/2,2,5/2,3$. For the uncoupled engine ($J=0$),  the engine efficiency can be calculated as 
$\eta_{J=0}=1-B_2/B_1$ which is independent of spin $s$ as can be seen in Fig.~\ref{fig:fig1}(b). The coupled engine performance can be higher than the uncoupled one; both $W$ and $\eta$ first increase to certain maximums as a function of $J$ and then drop to zero. 
The role of spin $s$ on the global work and efficiency is found to shift the maximums and the positive work conditions (PWCs) 
to the weak coupling regimes; accordingly the 
coupled Otto engine with high spin $s$ can produce higher work with higher efficiency than the lower spin $s$, below a certain sufficiently weak coupling strength (for instance,  $J<\approx 0.12$ in Fig.~\ref{fig:fig1}). 
Especially, the engine with $s>1/2$ can violate $\eta_b$ as indicated by the dashed line in Fig.~\ref{fig:fig1}(b).

\begin{figure}[!ht]\centering
\includegraphics[width=8.5cm]{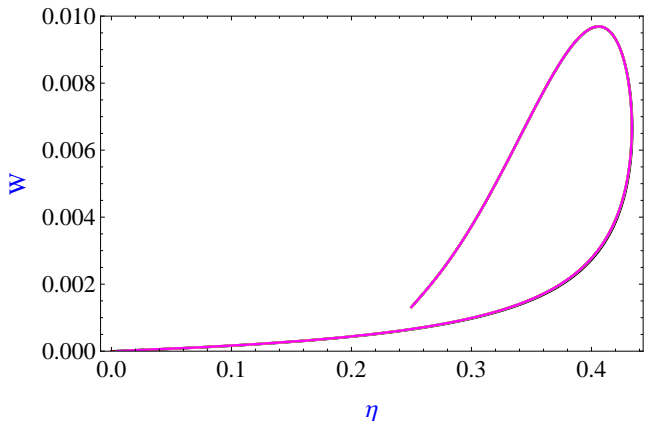}
\caption{\label{fig:fig2} (Color online.) Mutual relation of global work $W$ and efficiency $\eta$ for the same magnetic field and temperature parameters and the coupling range as in Fig.~\ref{fig:fig1}. The curves for each spin $s$ nearly coincide.}
\end{figure}
The mutual relationship between the work output and efficiency is demonstrated by the characteristic curve in Fig.~\ref{fig:fig2},
for the same magnetic field and temperature values, and for the same coupling strength range as in Fig.~\ref{fig:fig1}.
It can be deduced from Fig.~\ref{fig:fig2} that the efficiency at maximum work output as well as the work at maximum efficiency are 
not notably affected by the spin $s$ of the working substance. It seems that the higher spin $s$ leads to higher efficiency 
and work output at the weak coupling regime. We should stress here that this is not the general conclusion; for differently 
tailored parameters, the maximum of work output and the efficiency can slightly be influenced by the spin $s$. 

\begin{figure}[!ht]\centering
\includegraphics[width=8.5cm]{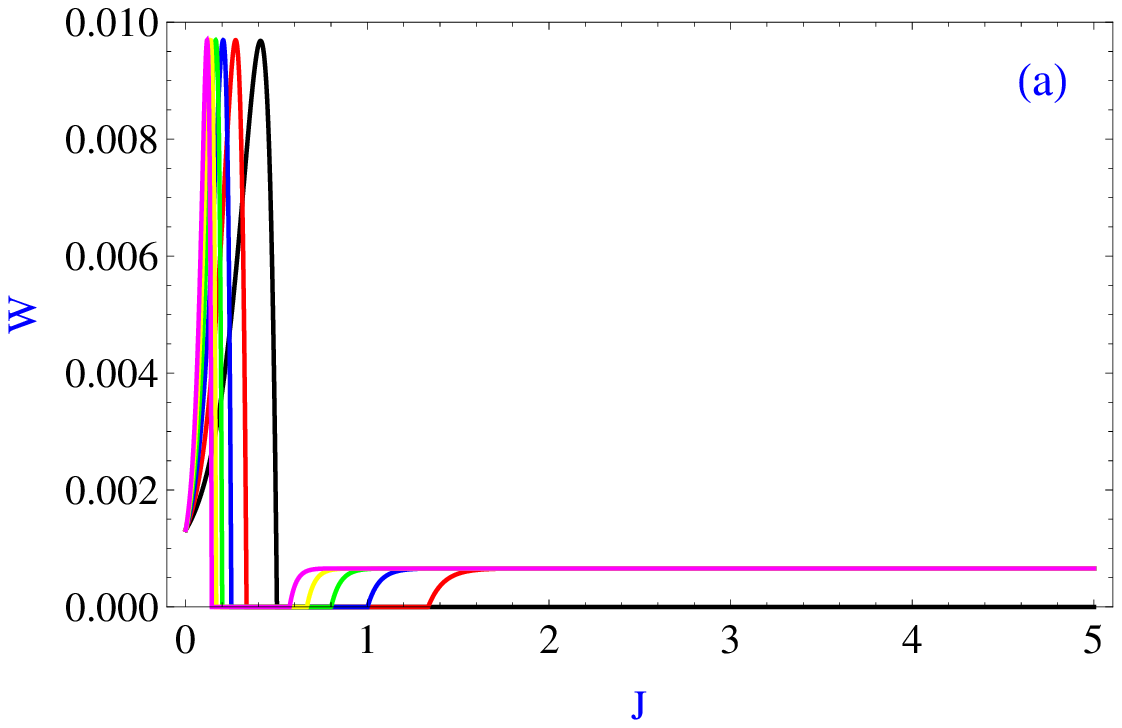}
\includegraphics[width=8.5cm]{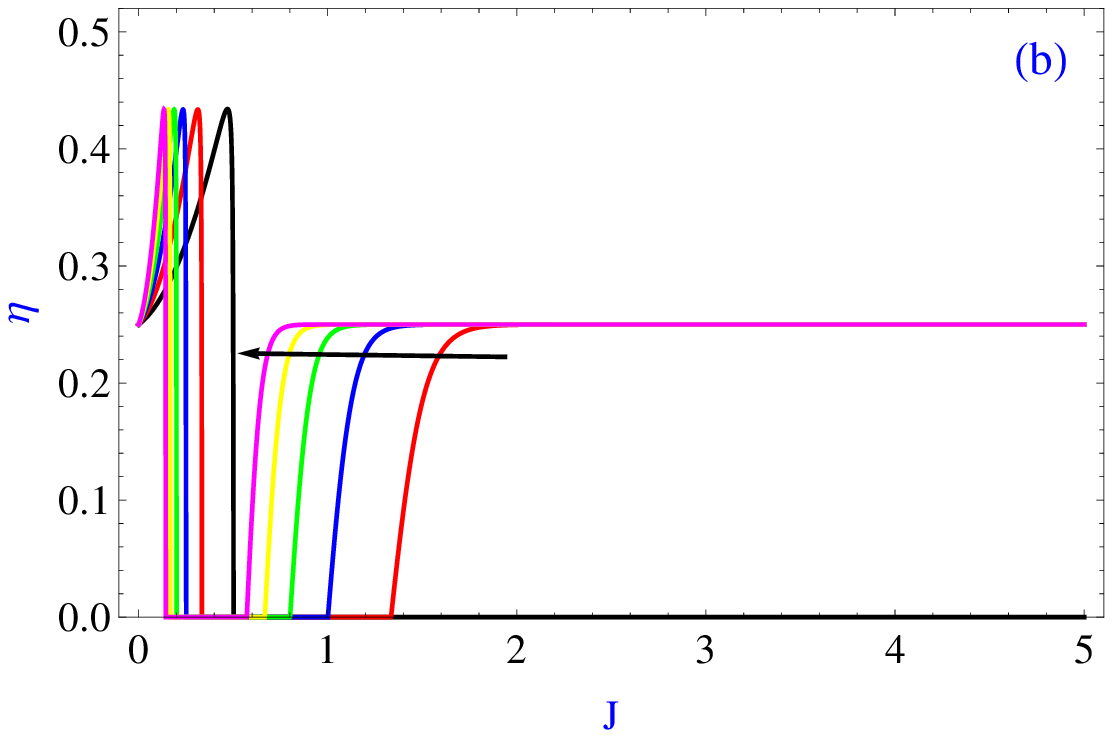}
\caption{\label{fig:fig3} (Color online.) The work output (a) and efficiency (b) in the broader range of $J$, including the strong coupling region for the same parameters and spin $s$ as in Fig.~\ref{fig:fig1}. The direction of arrow in (b) indicates the lines in the order of increasing spin $s$ from $s=1$ to $s=3$. The curves are the same with those in Fig.~\ref{fig:fig1} in the weak coupling regime. Note that after $J\approx 0.5$, the engine cannot produce positive work for the case of coupled spins $1/2$.}
\end{figure}

In Fig.~\ref{fig:fig1}, we have restricted ourselves to the weak coupling regime, specifically $J\in[0,0.5]$, and now we focus on the strong coupling region. It is possible to show that beyond this limit, i.e., $J>0.5$, the working substance of two spins $1/2$ cannot do positive work, since it violates the PWC given in Ref.~\cite{thomas11}. It is reasonable to assume that the change of energy gaps in the adiabatic stages by the change of magnetic field cannot contribute in the direction of total positive work gradient when $J>0.5$. On the other hand, for the case of pairing spin $1/2$ and spin $s$ with $s>1/2$, the role of energy gaps in the work extraction can be dramatically changed after a critical value of $J$ and the engine can reproduce useful work. This is shown in Fig.~\ref{fig:fig3} where the global work and efficiency are plotted as a function of $J$ up to the very strong couplings. As shown in Fig.~\ref{fig:fig3}, the positive work re-emerges after a critical value of coupling strength. Increasing the spin $s$ value shifts the critical $J$ towards the weak coupling regime.  The efficiency and the work output are less in the strong coupling regime. Since the corresponding thermodynamical quantities are invariant under uniform energy shifts~\cite{quan07}, the coupled spin $1/2$ and spin $s$ model in the limit of very large coupling strengths (i.e., $J\rightarrow\infty$) can be mapped into a multilevel system with energy levels $\{0,2B,4B,\ldots,(2s-1)2B\}$ where $\eta=0$ for $s=1/2$, while $\eta=1-B_2/B_1$ for $s>1/2$. This explains the behavior of the efficiency in Fig.~\ref{fig:fig3}(b) where $\eta$ converges to the
spin independent value of $\eta=1-B_2/B_1$ for $s>1/2$ and $\eta=0$ for $s=1/2$ in the deep strong coupling regime.
\section{Local Work and Efficiency}\label{sec:LocalWorkEfficiency}
In this section, we investigate how the spin $1/2$ and spin $s$ individually undergo the engine operation. This can be done by the analysis of local heat exchanges between the local spin and the reservoir~\cite{thomas11}. The local heat exchanges in the isochoric branches of the Otto cycle can be expressed as the change in the local density matrix for a given local Hamiltonian. Let $q_1^i$ ($q_2^i$), with $i=A,B$, 
be the local heat transferred between the $i$th spin and the hot (cold) heat bath. Then the explicit expression of $q_1^i$ ($q_2^i$) reads as~\cite{thomas11}:
\begin{eqnarray}\label{heatlocal}
q_1^i&=&Tr[(\rho_i-\rho_i^{'})H_i],\nonumber\\
q_2^i&=&Tr[(\rho_i^{'}-\rho_i)H_i^{'}],
\end{eqnarray}
where $\rho_i$ $(\rho_i^{'})$ is the reduced density matrix for the $i$th spin at the end of stage 1 (3) and $H_i$ ($H_i^{'}$) is the local Hamiltonian during the first (second) isochoric process. The local Hamiltonians can be written as $H_A=2Bs_A^z$ and $H_B=2BS_B^z$ for the spin $1/2$ and spin $s$, respectively. The local work done by the $i$th spin is then written as $w_i=q_1^i+q_2^i$.

\begin{figure}[!ht]\centering
\includegraphics[width=8cm]{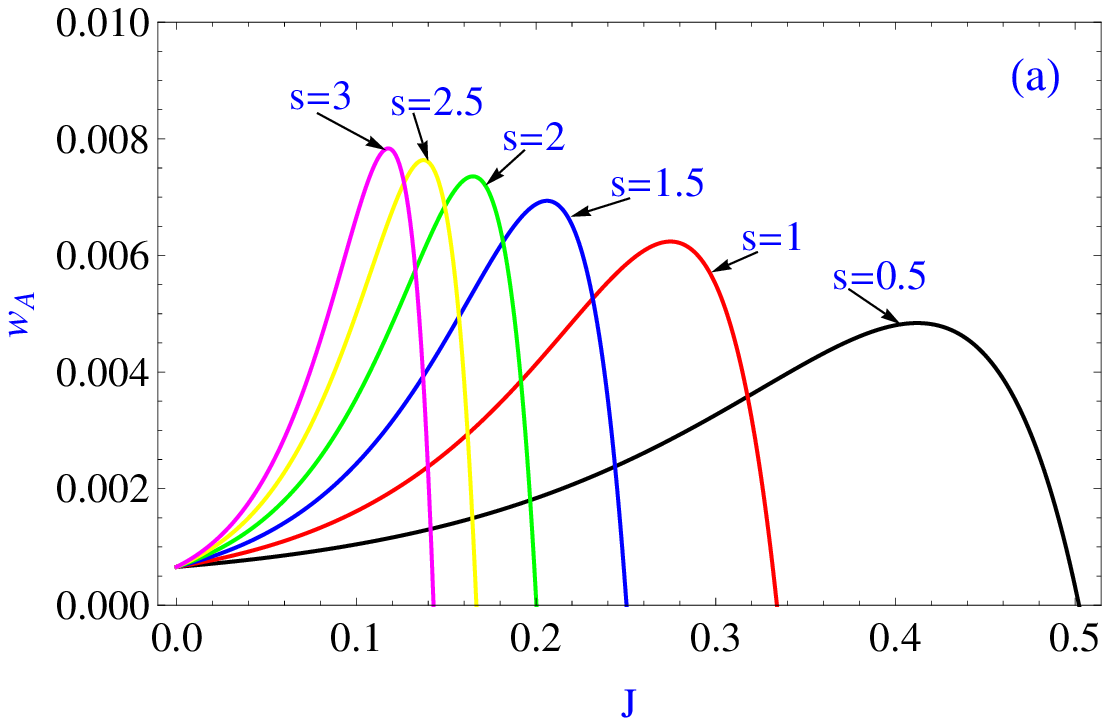}
\includegraphics[width=8cm]{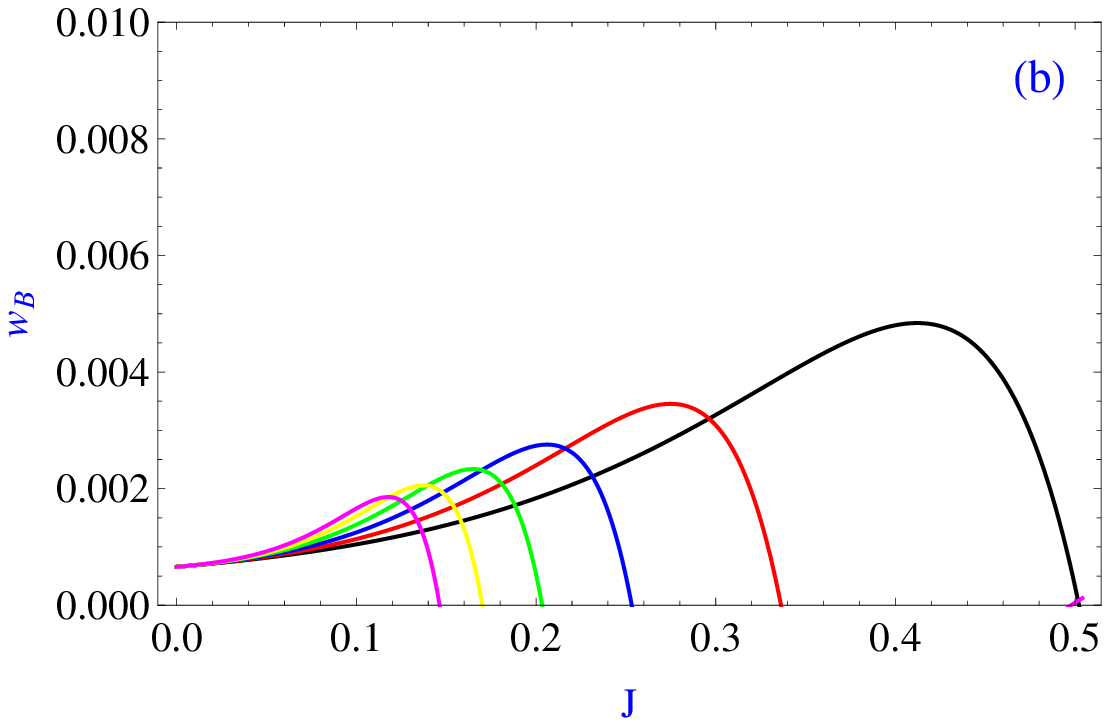}
\caption{\label{fig:fig4}(Color online.) The local work done by the spin $1/2$ (a) and spin $s$ (b) versus coupling strength $J$ for values $T_1=1$, $T_2=0.5$, $B_1=4$, $B_2=3$, and $s=1/2$~(black line), $s=1$~(red line), $s=3/2$~(blue line), $s=2$~(green line), $s=5/2$~(yellow line) and $s=3$~(magenta line).}
\end{figure}
The local works $w_A$ and $w_B$, done by the spin $1/2$ and spin $s$, respectively, are plotted as a function of 
coupling strength $J$ in Fig.~\ref{fig:fig4} for $s=1/2,1,3/2,2,5/2,3$. The analytical calculation of the global and local works yields that $W=w_A+w_B$, the total work is the sum of local efforts. 
For further insight, it is possible to calculate the relation between the global and local heat exchanges, which is found to be 
\begin{eqnarray}\label{heatlocalglobal}
Q_1&=&q_1^A+q_1^B+8J \mathcal{P}_s,\nonumber\\
Q_2&=&q_2^A+q_2^B-8J \mathcal{P}_s,
\end{eqnarray}
where  $\mathcal{P}_s=Tr[(\rho-\rho^{'})\vec{s}_A.\vec{S}_B]$, $\rho$~$(\rho^{'})$ being the global thermal density matrix at the end of stage 1 (3) given by Eq.~(\ref{denmatAB}).  The term $\mathcal{P}_s$, given by expectation value of interacting part, is related to the probabilities of certain energy levels at the end of stages $1$ and $3$. Its explicit expression depends on the spin-$s$ but not written here explicitly for brevity. The relations in Eq.~(\ref{heatlocalglobal}) suggest that only the local heat exchange is converted into total work output of the Otto cycle, as the last terms in $Q_1$ and $Q_2$ expressions reflect the collective heat intake and release which cancel each other. This is consistent with the extensive property of the work output of the cycle. We should stress here that same conclusion is reached for the case of spin $1/2$~\cite{thomas11} and spin $3/2$ pairs~\cite{ivanchenko14}. The extensive property is not a fundamental character of the work output and is not always true. Similar analysis in different conditions reveal that sum of the local works is not always equal to the global work~\cite{huang14plus,huang13}. We will present a more general discussion in the following section.

\begin{figure}[!ht]\centering
\includegraphics[width=8cm]{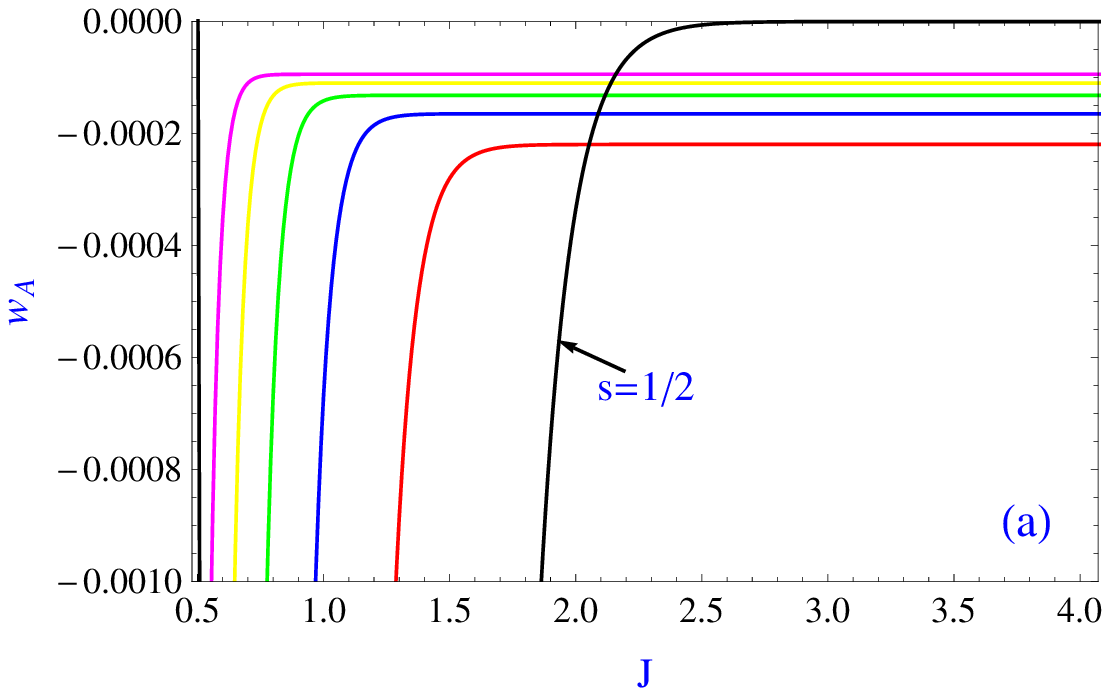}
\includegraphics[width=8cm]{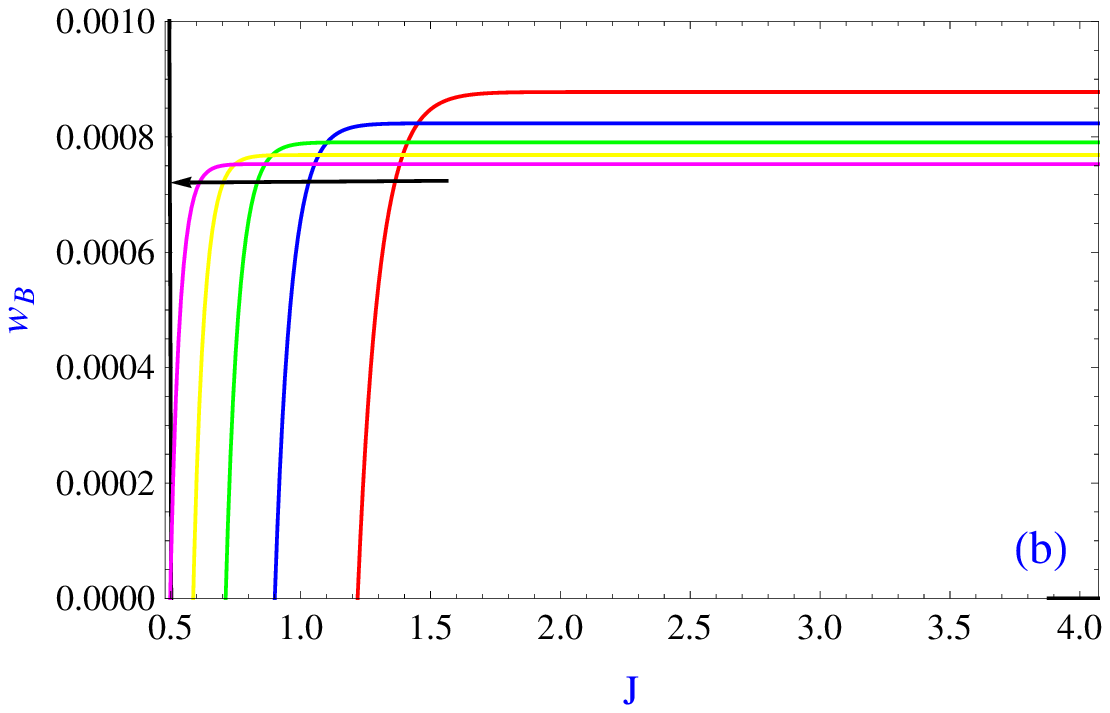}
\caption{\label{fig:fig5} (Color online.) The local work done by the spin $1/2$ (a) and spin $s$ (b) in the strong coupling region where $J>0.5$ for the same parameters and spin $s$ as in Fig.~\ref{fig:fig4}. The curves are the same with those in Fig.~\ref{fig:fig4} in the weak coupling regime. The direction of arrow in (b) indicates the lines in the order of increasing spin $s$ from $s=1$ to $s=3$. Note that after $J\approx 0.5$, $w_A=w_B\leq 0$ for $s=1/2$ as given black line in (a).}
\end{figure}

For two coupled spin $1/2$ case, we have $w_A=w_B$ since $\rho_A=\rho_B$ and $H_A=H_B$~\cite{thomas11}. Moreover, for $J=0$, 
$w_A$ is independent of spin $s$ value. $w_B$ depends on spin $s$ for $J=0$, but this dependence is weak
to be visible in the scale of Fig.~\ref{fig:fig4}. 
On the other hand, these results are dramatically changed when $s>1/2$ and $J\neq 0$. 
As shown in Fig.~\ref{fig:fig4}(a), $w_A$ depends strongly on the spin $s$. In the region $J<0.5$, increasing $s$ shifts the PWCs and maximums of $w_A$ and $w_B$ to the weak coupling regions and increases (decreases) the maximums of $w_A$ ($w_B$). The comparison of local works of both spins shows that, except a negligibly tiny range of $J$, we have $w_A>w_B$, that is spin $1/2$ does more work than the spin $s$. On the other hand, if we change our attention to the strong coupling regime where $J>0.5$ (Fig.~\ref{fig:fig5}), this situation is completely reversed; as shown in Fig.~\ref{fig:fig5}(a), $w_A\leq 0$ for each spin $s$, while $w_B$ can be non-zero for $s>1/2$ (Fig.~\ref{fig:fig5}(b)). From an analytical calculation of global and local works in the deep strong coupling regime (i.e., $J\rightarrow\infty$), it is possible to show that $W=-(2s+1)w_A=(2s+1)/(2s+2)w_B$. This indicates that spin $s$ is solely responsible for the realization of our QHE in the strong coupling regime, where $w_B>0$ and $w_A<0$ in the regions $W>0$. 

 For further insight of the local performance, we analyze the local efficiency and the local heat flow of spin $1/2$ and spin $s$. From the local work and heat exchange definitions in Eq.~(\ref{heatlocal}), for the considered case (i.e., $B_1>B_2>0$~\cite{exp}), we always have $q_1^i>-q_2^i>0$ when $w_i>0$; that is local heat always flow in the direction of global heat gradient when the local work is positive. For two spins $1/2$, local spins are always heat engines when the global work $W>0$, since $w_A=w_B=w$, $W=2w$ and $W,w>0$. On the other hand, this is not necessarily true for the two coupled asymmetric spins. One of the spins can be refrigerant when total 
system operates as a QHE. From the comparison of Fig.~\ref{fig:fig1} and Fig.~\ref{fig:fig4}, one can deduce that there is a critical $J$ depending on $s$ up to which the local spins are heat engines. Beyond the critical coupling strength, as can be seen in Fig.~\ref{fig:fig5}, we have $w_B>0$ and $w_A<0$ ($q_1^A<0$ and $q_2^A>0$) when $W>0$. Here spin $1/2$ acts in the opposite direction of global work gradient and it is a refrigerator, although total system and the spin $s>1/2$ are the heat engines. 

For the local heat exchanges in Eq.~(\ref{heatlocal}), we have the relation $q_1^i=-(B_1/B_2)q_2^i$, so  we found that the individual spins undergo the heat cycle with the same and constant local efficiency: $\eta_A=\eta_B=w_i/q_1^i=1-B_2/B_1$, which is independent of spin $s$ and equals to the global uncoupled engine efficiency. In the regime where spin $1/2$ operates locally in the refrigeration cycle it
is more appropriate to consider local coefficient of performance, which is also a constant $\epsilon_A=q_2^A/|w_A|=B_2/(B_1-B_2)$.

\begin{figure}[!ht]\centering
\includegraphics[width=8cm]{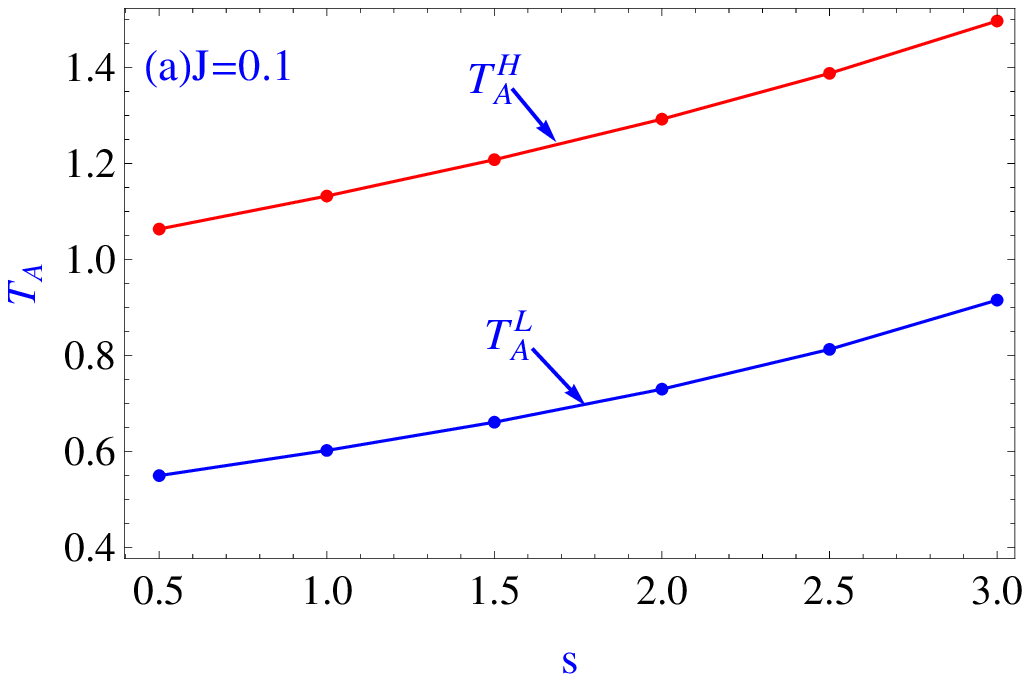}
\includegraphics[width=8cm]{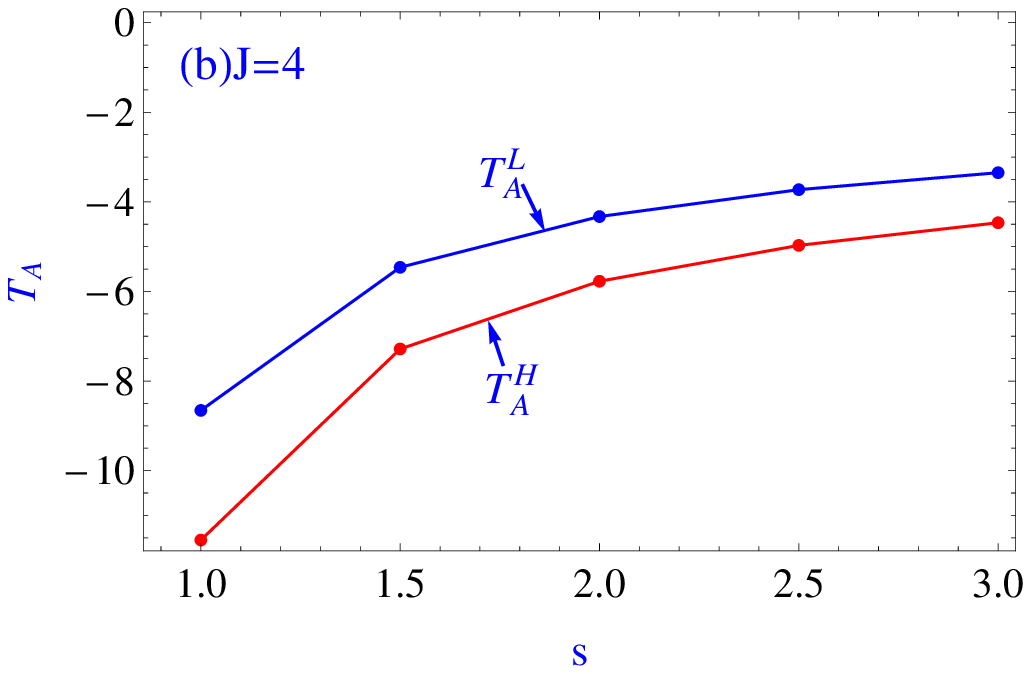}
\caption{\label{fig:fig6}(Color online.) The local temperature of spin $1/2$ versus the parameter $s$ at $J=0.1$ (a) and at $J=4.0$ (b). Here $T_A^H$ $(T_A^L)$ denotes the local temperature for the hot (cold) heat bath cases with $T_1=1.0$ $(T_2=0.5)$ and $B_1=4$ $(B_2=3)$.}
\end{figure}

 Now we focus on the temperatures of local spins at the end of two thermalization stages and investigate the role of coupling $J$ and spin $s$ on the local temperatures. By taking two different energy levels, $E_i$, and their probabilities, $P_i$ (obtained from the reduced density matrix in Eq.~(\ref{heatlocal})), one can define an effective temperature for the local spins $A$ and $B$ as:
\begin{eqnarray}\label{localtemp}
T_k=\frac{E_i-E_j}{\ln P_j-\ln P_i}, \quad k=A,B.
\end{eqnarray}
In the absence of coupling between spins (i.e., $J=0$), the local temperature of both spins are equal to the heat bath temperature, since we have $\rho=\rho_A\otimes\rho_B$, where $\rho_{i}~(i=A,B)$ are the local thermal density matrices at the heat bath temperature. On the other hand, this separation would not be possible in the case of nonzero $J$ which can make the spins to locally thermalize to different temperatures than the heat bath~\cite{thomas11}. Analysis of the local temperature of spin $s>1/2$ reveals that the reduced state of spin $s$ is a non-equilibrium steady state; where there is no unique effective temperature applicable to all the energy level pairs~\cite{quan07}. For the two-level system (spin $1/2$) it would always be possible to define an effective temperature. 

Interplay of interactions and multilevel nature of large spin on the thermalization can be explained at more intuitive level as follows. Our model can be interpreted as a central spin coupled to an ensemble of $2s$ spins $1/2$~\cite{bortz07,prokof00}. Both the collective spin bath and the central spin are also coupled to a heat reservoir. The heat bath and the spin $s>1/2$ can always fully thermalize the spin $1/2$; while spin $s>1/2$ cannot be thermalized when it is coupled to the spin $1/2$. This intuitive reasoning based upon the size difference of subsystems is in parallel with the conclusion in Ref.~\cite{klein15} which states that the minimum number of spins $1/2$ required to thermalize a single spin $1/2$ is two. 

The effective temperature $T_A$ of spin $1/2$ with respect to spin $s$ at the end of first ($T_A^H$) and second ($T_A^L$) isochoric processes for weak and ultrastrong coupling regimes are shown in Fig.~\ref{fig:fig6}(a) and in Fig.~\ref{fig:fig6}(b), respectively.
At weak coupling regime ($J=0.1$), where the spin $1/2$ is a heat engine as shown in Fig.~\ref{fig:fig4}(a), $T_A$ is always larger than the heat bath temperature. Spin $s$, acting as an additional reservoir next to the heat bath, can heat the central spin $1/2$ to higher temperatures. For two spins $1/2$, increasing $J$ makes the occupation probabilities of energy levels of reduced spins equal 
so that effective temperature becomes infinitely high at the ultrastrong coupling regime. On the other hand, for $s\geq 1$, after a critical coupling strength the population in high energy level ($E_{\uparrow}=B$) exceeds the population in lower energy level   ($E_{\downarrow}=-B$) so that local temperature of the spin $1/2$ becomes negative as shown in Fig.~\ref{fig:fig6}(b). After the first and the second adiabatic stages, the local temperatures of the spin $1/2$ change to $(B_2/B_1)T_A^H$ and $(B_1/B_2)T_A^L$, respectively. Fig.~\ref{fig:fig6}(a) indicates that $T_A^H>(B_1/B_2)T_A^L$ and $T_A^L<(B_2/B_1)T_A^H$, so that $q_1^A>0$ and $q_2^A<0$; hence  $w_A>0$ as discussed above. In the deep strong coupling regime, Fig.~\ref{fig:fig6}(b) elucidates that $T_A^H<(B_1/B_2)T_A^L$ and $T_A^L>(B_2/B_1)T_A^H$, so that $q_1^A<0$, $q_2^A>0$; consequently $w_A<0$, so that spin $1/2$ is a refrigerator.

\section{General relations between global and local work}\label{sec:generalGlobalLocalWork}
We have seen in Sec.~\ref{sec:LocalWorkEfficiency} that the global work has an extensive property
and can be written as a sum of the local works done by the individual spins. This conclusion strictly depends on the paths,
or the methods, we choose to operate the engine cycle.  In the adiabatic stages of the quantum Otto
cycle, we varied the homogeneous magnetic field acting on the spins. 
We can make a general statement that it is not possible break the extensive property of work output of a QHE by only
making local changes in the adiabatic stages of the engine cycle. This simple fact can be quickly proven for a general Hamiltonian
of a system of a collection of local subsystems, described in the form $H=\sum H_{\mathrm{loc}}+H_{\mathrm{int}}$, 
where the non-interacting (local) and interacting (global) terms are denoted
by $H_{\mathrm{loc}}$ and $H_{\mathrm{int}}$, respectively. The internal energy, $U=\langle H\rangle=\mathrm{Tr}(\rho H)$, 
of the system with density matrix $\rho$ changes as $dU=\mathrm{Tr}(\rho dH)+\mathrm{Tr}(Hd\rho)$, where the first term
can be defined as the work done on the system and denoted by $\dbar W:=\mathrm{Tr}(\rho dH)$.
In a strictly quantum adiabatic process we have $d\rho=0$. Accordingly, if $dH_{\mathrm{int}}=0$, 
the global work becomes extensive in terms of local works done by subsystems such that 
$\dbar W=\sum \dbar w_{\mathrm{loc}}$, with 
$\dbar w_{\mathrm{loc}}:=\mathrm{Tr}_{\mathrm{loc}}(\rho_{\mathrm{loc}} dH_{\mathrm{loc}})$, 
where $\rho_{\mathrm{loc}}$ is the reduced
density matrix of a particular subsystem found by tracing out the degrees of freedom of the other subsystems from the density matrix
$\rho$ of the whole system. While the global work is extensive under local changes, it can still be optimized by the interactions
between the subsystems, through the interaction dependence of the reduced density matrices  $\rho_{\mathrm{loc}}$, which is
illustrated by our analysis in Sec.~\ref{sec:globalWorkEfficiency} and Sec.~\ref{sec:LocalWorkEfficiency}.

Let us now consider a more general situation where
both the magnetic field and the exchange interaction between the spins could change. In such a case, Eq.~(\ref{heatlocalglobal})
directly shows that the extensive behavior of the global work is violated by the simultaneous change of magnetic field strength
($B_1\rightarrow B_2\rightarrow B_1$) and the exchange coupling strength ($J_1\rightarrow J_2\rightarrow J_1$) in the adiabatic
stages such that
\begin{eqnarray}\label{eq:generalGlobalLocalW}
W=w_A+w_B+8(J_1-J_2)\mathcal{P}_s,
\end{eqnarray}
where  $\mathcal{P}_s$ is defined in Eq.~(\ref{heatlocalglobal}). 

A curious result of Eq.~(\ref{eq:generalGlobalLocalW}) is that when $B_1=B_2$ and $J_1\neq J_2$, the system can harvest positive work
in a purely collective manner, as no local work can be done by the local systems in constant magnetic field. Since there is no change in
local Hamiltonians, the total local heat exchange is zero.  If we take the ratio $W/w_{\mathrm{loc}}$, where $w_{\mathrm{loc}}=w_A+w_B$ is the total local work, as a figure of merit measuring the cooperativity in work extraction, it is infinite. On the other hand,
we can still consider a possible generalization of the local work definition in Ref.~\cite{thomas11} to scrutiny them in a purely interacting cycle without explicit local variations. We suggest that a mean field Hamiltonian can always be introduced to describe a local Hamiltonian of a subsystem.

To make our discussion concrete let us take a pairwise interaction Hamiltonian of the form $H=gAB$, where $A$ and $B$ are operators for two
subsystems, and $g$ is their coupling constant. The work done on the system in an adiabatic stage by the $dg$ variation of the coupling constant can be written as $\dbar W = dg\langle AB\rangle$, where $\langle AB\rangle=\mathrm{Tr}(\rho AB)$. If we use mean field Hamiltonians $H_A=g\langle B\rangle A/2$ and $H_B=g\langle A\rangle B/2$ for the local Hamiltonians then the corresponding local work contributions
become $w_A=w_B=g\langle A\rangle\langle B\rangle/2$. Accordingly the global work can be expressed as 
$\dbar W=\dbar w_A+\dbar w_B+\dbar w_{\mathrm{coop}}$, where we introduced a cooperative work term 
$\dbar w_{\mathrm{coop}}:=dg\langle A,B\rangle$. Here, the
notation $\langle A,B\rangle:=\langle AB\rangle-\langle A\rangle\langle B\rangle$ stands for the covariance of $A$ and $B$ as a measure of
correlations between the subsystems. The net work done in the cycle then becomes 
\begin{eqnarray}
W=w_A+w_B+w_{\mathrm{coop}},
\end{eqnarray}
where the local and cooperative works are given by 
\begin{eqnarray}
w_A&=&w_B=\frac{1}{2}(g_1-g_2)\left(\langle A\rangle_1\langle B\rangle_1
-\langle A\rangle_2\langle B\rangle_2\right),\nonumber\\
w_{\mathrm{coop}}&=&(g_1-g_2)(\langle A,B\rangle_1-\langle A,B\rangle_2).
\end{eqnarray}
Here $g_1$ and $g_2$ are the coupling constants at the end points of the adiabatic stages, and the expectation values
$\langle X\rangle_i=\mathrm{Tr}(\rho_X^iX)$ are evaluated with the reduced density matrix $\rho_X^i$ of the subsystem $X=A,B$
in the adiabatic stage labeled by $i=1,2$. With this generalized definition of the local work, the cooperativity of the work extraction
can be characterized by the ratio
\begin{eqnarray}
\frac{W}{w_{\mathrm{loc}}}=1+\frac{\langle A,B\rangle_1-\langle A,B\rangle_2}{\langle A\rangle_1\langle B\rangle_1
-\langle A\rangle_2\langle B\rangle_2}.
\end{eqnarray}

Applying the generalized local work formalism to our Heisenberg exchange model QHE, we find the local Hamiltonians
\begin{eqnarray}
H_A&=&2Bs_A^z+\frac{1}{2}8J\vec{s}_A.\langle \vec{S}_B\rangle,\nonumber\\
H_B&=&2BS_B^z+\frac{1}{2}8J\langle\vec{s}_A\rangle. \vec{S}_B,
\end{eqnarray}
which gives the relation between global and local works as $\dbar W=\dbar w_A+\dbar w_B+\dbar w_{\mathrm{coop}}$, where
\begin{eqnarray}
\dbar w_A&=&2dB\langle s_A^z\rangle+\frac{1}{2}8dJ\langle\vec{s}_A\rangle.\langle \vec{S}_B\rangle,\nonumber\\
\dbar w_B&=&2dB\langle S_B^z\rangle+\frac{1}{2}8dJ\langle\vec{s}_A\rangle. \langle \vec{S}_B\rangle,
\end{eqnarray}
and $\dbar w_{\mathrm{coop}}=8dJ\langle\vec{s}_A, \vec{S}_B\rangle$. From this result we conclude that the extensive property
of the global work can be violated by changing the interaction parameter in the adiabatic stages, if the covariance of the interacting
spins changes as well. If the covariance remains the same, then the global work can be expressed as the sum of effective local works
of the individual spins under the mean field description.

\section{Conclusions}\label{sec:conclusion}
We consider a pair of spin $1/2$ and spin $s$ coupled via Heisenberg exchange interaction under a homogeneous magnetic field 
as the working medium of a quantum Otto engine. The influence of exchange coupling and spin $s$ on the work output and 
efficiency of the quantum Otto engine is investigated in detail. The global engine operation is also analyzed in comparison to local work 
contributions of the individual spins. It is found that increasing spin $s$ at a certain exchange coupling strength can make the 
QHE to produce more work with higher efficiency, which can violate the upper bound of efficiency for two coupled spin $1/2$ particles~\cite{thomas11}. Moreover, spin $s$ makes it possible to realize the QHE at the strong coupling regimes.  Furthermore, we show that due to the coupling of asymmetric spins, one of the spins can operate as a refrigerator even when global cycle is a heat engine. From the local work analysis, it is found that global work is equal to the sum of the local works by the individual spins. Although in local realm, the spin $1/2$ and spin $s$ operate with the same efficiency, their local works are found to be 
significantly influenced by the spin $s$.  The local temperature of spin $1/2$ is found to be controlled by spin $s$ and can be negative or larger than the temperature of the heat baths in the case of non-zero coupling.  Finally we discussed the conditions for the violation of the extensive behavior of the global
work. We developed a formalism, applicable to any coupled bipartite system, generalizing the local and cooperative work 
definitions to the case where global changes can be performed in the engine cycle. The general conditions for which the global work is not equal to the sum of the local works are given in terms of the covariance of the subsystems.

\acknowledgments
F.~A. thanks R.~Eryigit, G.~Thomas, and R.~S.~Johal for fruitful discussions. 
\"O.~E.~M. acknowledges illuminating comments by N.~Allen and support from Ko\c{c} University and Lockheed Martin
University Research Agreement.
\\
 \appendix*
\section{The Eigenvalues of the Working Medium}\label{appendix:eigenvalues}
Here we report the eigenvalues of the Hamiltonian~(\ref{hamiltonian}) for $s=1/2,1,3/2,2,5/2,3$. The corresponding orthonormal eigenstates can also be calculated. We should stress here that the eigenstates are system parameter (i.e., $J$ and $B$) independent. Since the discussion of text does not require the explicit form of the eigenstates, we do not report them here for brevity.

The eigenvalues for $\left(\frac{1}{2},s\right)$ system with $s=1/2$ are~\cite{thomas11}: $\{-6J,2J-2B,2J,2J+2B\}$.

The eigenvalues for $\left(\frac{1}{2},s\right)$ system with $s=1$ are: $\{-B-8J,B-8J,-3B+4J,-B+4J,B+4J,3B+4J\}$.

The eigenvalues for $\left(\frac{1}{2},s\right)$ system with $s=3/2$ are: $\{-2B-10J,-10J,2B-10J,-2B+6J,-4B+6J,6J,2B+6J,4B+6J\}$.

The eigenvalues for $\left(\frac{1}{2},s\right)$ system with $s=2$ are: $\{-3B-12J,-B-12J,B-12J,3B-12J,-5B+8J,-3B+8J,-B+8J,B+8J,3B+8J,5B+8J\}$.

The eigenvalues for $\left(\frac{1}{2},s\right)$ system with $s=5/2$ are: $\{-4B-14J,-2B-14J,-14J,2B-14J,4B-14J,-2B+10J,-4B+10J,-6B+10J,10J,2B+10J,4B+10J,6B+10J\}$.

The eigenvalues for $\left(\frac{1}{2},s\right)$ system with $s=3$ are: $\{-5B-16J,-3B-16J,-B-16J,B-16J,3B-16J,5B-16J,-3B+12J,3B+12J,-7B+12J,-5B+12J,-B+12J,B+12J,5B+12J,7B+12J\}$.


\begin{thebibliography}{99}
\bibitem{scovil59} H.E.D. Scovil and E.O. Schulz-DuBois, Phys. Rev. Lett. \textbf{2}, 262 (1959).

\bibitem{quan07} H.T. Quan, Yu-xi Liu, C.P. Sun and F. Nori, Phys. Rev. E \textbf{76}, 031105 (2007).

\bibitem{quan09} H.T. Quan, Phys. Rev. E \textbf{79}, 041129 (2009).

\bibitem{kieu04} T.D. Kieu, Phys. Rev. Lett. \textbf{93}, 140403 (2004).

\bibitem{jwang12} J. Wang, Z. Wu and J. He, Phys. Rev. E \textbf{85}, 041148 (2012).

\bibitem{wang13} R. Wang, J. Wang, J. He and Y. Ma, Phys. Rev. E \textbf{87}, 042119 (2013).

\bibitem{uzdin14} R. Uzdin and R. Kosloff, EPL \textbf{108}, 40001 (2014).

\bibitem{quan05} H.T. Quan, P. Zhang and C.P. Sun, Phys. Rev. E \textbf{72}, 056110 (2005).

\bibitem{allahverdyan08} A.E. Allahverdyan, R.S. Johal and G. Mahler, Phys. Rev. E \textbf{77},  041118 (2008).

\bibitem{altintas14} F. Altintas, A.U.C. Hardal and O.E. Mustecaplioglu, Phys. Rev. E \textbf{90}, 032102 (2014).

\bibitem{thomas11}  G. Thomas and R.S. Johal, Phys. Rev. E \textbf{83}, 031135 (2011).

\bibitem{zhang08} G.-F. Zhang, Eur. Phys. J. D \textbf{49}, 123-128 (2008).

\bibitem{huang13} X.L. Huang, L.C. Wang and X.X. Yi, Phys. Rev. E \textbf{87}, 012144 (2013).

\bibitem{feldmann04} T. Feldmann and R. Kosloff, Phys. Rev. E \textbf{70}, 046110 (2004).

\bibitem{feldmann03} T. Feldmann and R. Kosloff, Phys. Rev. E \textbf{68}, 016101 (2003).

\bibitem{kosloff02}  R. Kosloff and T. Feldmann, Phys. Rev. E \textbf{65}, 055102 (2002).

\bibitem{henrich07} M.J. Henrich, G. Mahler and M. Michel, Phys. Rev. E \textbf{75}, 051118 (2007).

\bibitem{zhang07} T. Zhang, W.-T. Liu, P.-X. Chen and C.-Z. Li, Phys. Rev. A \textbf{75}, 062102 (2007).

\bibitem{thomas14} G. Thomas and R.S. Johal, Eur. Phys. J. B  \textbf{87}, 166 (2014).

\bibitem{huang12} X.L. Huang, T. Wang and X.X. Yi, Phys. Rev. E \textbf{86}, 051105 (2012).

\bibitem{huang14} X.-L. Huang, X.-Y. Niu, X.-M. Xiu and X.-X. Yi, Eur. Phys. J. D  \textbf{68}, 32 (2014).

\bibitem{wu06} F. Wu, L. Chen, F. Sun, C. Wu and Q. Li, Phys. Rev. E \textbf{73}, 016103 (2006).

\bibitem{ivanchenko14} E.A. Ivanchenko, arXiv:1412.7171.

\bibitem{wang12} H. Wang, G. Wu and D. Chen, Phys. Scr. \textbf{86}, 015001 (2012).

\bibitem{he12} X. He, J. He and J. Zheng, Physica A \textbf{391}, 6594 (2012).

\bibitem{huang14plus} X.L. Huang, Y. Liu, Z. Wang and X.Y. Niu, Eur. Phys. J. Plus \textbf{129}, 4 (2014).

\bibitem{wang09} H. Wang, S. Liu and J. He, Phys. Rev. E \textbf{79}, 041113 (2009).

\bibitem{hübner14} W. Hubner, G. Lefkidis, C.D. Dong, D. Chaudhuri, L. Chotorlishvili and J. Berakdar, Phys. Rev. B \textbf{90}, 024401 (2014).

\bibitem{azimi14} M. Azimi, L. Chotorlishvili, S.K. Mishra, T. Vekua, W. Hubner and J. Berakdar,  New J. Phys. \textbf{16}, 063018 (2014).

\bibitem{albayrak13} E. Albayrak, Int. J. Quantum Inform. \textbf{11}, 1350021 (2013).

\bibitem{dillen09} R. Dillenschneider and E. Lutz, EPL \textbf{88}, 50003 (2009).

\bibitem{rezek06} Y. Rezek and R. Kosloff, New J. Phys. \textbf{8}, 83 (2006).

\bibitem{robnagel14} J. Robnagel, O. Abah, F. Schmidt-Kaler, K. Singer and E. Lutz, Phys. Rev. Lett. \textbf{112}, 030602 (2014).

\bibitem{zhang14} X.Y. Zhang, X.L. Huang and X.X. Yi, J. Phys. A: Math. Theor. \textbf{47}, 455002 (2014).

\bibitem{scully03} M.O. Scully, M.S. Zubairy, G.S. Agarwal and H. Walther, Science \textbf{299}, 862 (2003).

\bibitem{quan06} H.T. Quan, P. Zhang and C.P. Sun, Phys. Rev. E \textbf{73}, 036122 (2006).

\bibitem{altintas15} F. Altintas, A.U.C. Hardal and O.E. Mustecaplioglu, Phys. Rev. A \textbf{91}, 023816 (2015).

\bibitem{sothmann12} B. Sothmann and M. Buttiker, EPL \textbf{99}, 27001 (2012).

\bibitem{sothmann14} B. Sothmann, R. Sanches and A.N. Jordan, EPL \textbf{107}, 47003 (2014).

\bibitem{fialko12} P. Fialko and D.W. Hallwood, Phys. Rev. Lett. \textbf{108}, 085303 (2012).

\bibitem{zhangprl14} K. Zhang, F. Bariani and P. Meystre, Phys. Rev. Lett. \textbf{112}, 150602 (2014).

\bibitem{abah12} O. Abah, J. Robnagel, G. Jacob, S. Deffner, F. Schmidt-Kaler, K. Singer and E. Lutz, Phys. Rev. Lett. \textbf{109}, 203006 
(2012).

\bibitem{roncaglia14} A.J. Roncaglia, F. Cerisola and J.P. Paz, Phys. Rev. Lett. \textbf{113}, 250601 (2014);G.D. Chiara, A.J. Roncaglia and J.P. Paz, arXiv:1412.6116.

\bibitem{sinha03} S. Sinha, J. Emerson, N. Boulant, E.M. Fortunato, T.F. Havel, and D.G. Cory, Quantum Information Processing \textbf{2}, 433 (2003).

\bibitem{bortz07} M. Bortz and J. Stolze, J. Stat. Mech. p. P06018 (2007).

\bibitem{prokof00} N.V. Prokofev and P.C.E. Stamp, Rep. Prog. Phys. \textbf{63}, 669 (2000).

\bibitem{klein15} S. Kleinbolting and R. Klesse, Phys. Rev. E \textbf{91}, 052101 (2015).

\bibitem{li12} S.-S. Li, T.-Q. Ren, X.-M. Kong and K.Liu, Physica A \textbf{391}, 35 (2012).

\bibitem{exp}  For the case $B_2>B_1$, the local heats can flow in the direction opposite to the global heat gradient, i.e., $q_2^i>-q_1^i>0$ can be possible even when $Q_1>-Q_2>0$. Both local spins and global engine can produce positive work when $J\neq 0$. See Ref.~\cite{thomas11} for details.


\end{thebibliography}
\end{document}